\documentclass[showpacs,preprintnumbers,english,twocolumn,amsmath,amssymb]{revtex4}
\usepackage[T1]{fontenc}
\usepackage[latin1]{inputenc}
\usepackage{graphicx}
\usepackage{amssymb}
\usepackage{verbatim}
\usepackage{epic,eepic}
\usepackage{babel}
\begin{document}
\title{Higher Order Corrections to Density and Temperature of  Fermions from Quantum Fluctuations}
\author{ Hua Zheng$^{a,b)}$ and Aldo Bonasera$^{a,c)}$}
\affiliation{
a)Cyclotron Institute, Texas A\&M University, College Station, TX 77843, USA;\\
b)Physics Department, Texas A\&M University, College Station, TX 77843, USA;\\
c)Laboratori Nazionali del Sud, INFN, via Santa Sofia, 62, 95123 Catania, Italy.}
\begin{abstract}
A novel method to determine the density and temperature of a
system based on quantum Fermionic fluctuations  is
generalized to the limit where the reached temperature T is large
compared to the Fermi energy $\varepsilon_f$. 
 Quadrupole and particle multiplicity fluctuations
relations are derived in terms of $\frac{T}{\varepsilon_f}$. The
relevant Fermi integrals are numerically solved for any values of
$\frac{T}{\varepsilon_f}$ and compared to the analytical
approximations. The classical limit is obtained, as expected, in
the limit of large temperatures and small densities. We propose
simple analytical formulas which reproduce the numerical results,
valid for all values of $\frac{T}{\varepsilon_f}$.  The entropy can also be easily derived 
from quantum fluctuations and give important insight for the behavior of the system near a phase transition. A comparison of the quantum entropy to
the entropy derived from the ratio of the number of deuterons to {\it neutrons} gives a very good agreement especially when the density of the system is very low.
\end{abstract}
\pacs{ 25.70.Pq,42.50.Lc, 64.70.Tg}
\maketitle
The availability of heavy-ion accelerators which
provide colliding nuclei from a few MeV/nucleon to GeV/nucleon
has fueled a field of
research loosely referred to as Nuclear Fragmentation. The
characteristics of the fragments produced depend on the beam
energy and the target-projectile combinations which can be
externally controlled \cite{1,2,10}. Fragmentation  experiments
could provide informations about the nuclear matter properties and
constrain its EOS \cite{Csernai}.
 Long ago, W. Bauer stressed the crucial influence of the Pauli blocking
in  the momentum distributions of nucleons emitted in heavy ion collisions near the Fermi energy \cite{bauer}.
We have recently proposed a method to estimate the density and temperature based on fluctuations estimated
from an event by event determination of fragments arising after
the energetic collision \cite{plb11}.  A similar approach has also
been applied to observe experimentally the quenching of fluctuations in a trapped
Fermi gas \cite{prl}. We go beyond the method of \cite{prl} by
including quadrupole fluctuations as well to have a direct
measurement of densities and temperatures for subatomic systems.
 In this paper, we extend the method to derive the entropy of the system and we show how to recover the classical
limit when the temperatures are large compared to the Fermi
energy.
\begin{figure}
\centering
\includegraphics[width=1.0\columnwidth]{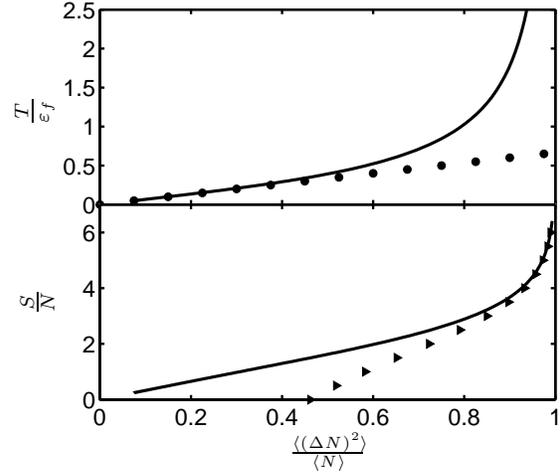}
\caption[]{ (Top) $\frac{T}{\varepsilon_f}$  versus multiplicity
fluctuations using different approximations. Full line gives the
numerical solution of Eq.  (\ref{numl_mul}), 
full dots are the lowest order approximation  discussed in ref.\cite{plb11};
(Bottom) Entropy per particle $\frac{S}{N}$ (in units of $\hbar$) versus multiplicity fluctuations.
Full line gives the numerical solution of Eq. (\ref{entropy}),
 full
triangles are the Sackur-Tetrod results.}\label{Fig1}
\end{figure}
We apply the proposed method to the microscopic CoMD approach
\cite{17} which includes Fermionic statistics. The resulting
energy densities and temperatures calculated using protons and
neutrons display a rapid increase around 3 MeV temperature which
is an indication of a first order phase transition.  This result is confirmed by the rapid increase of the entropy per unit volume
in the same temperature region. Similar results are found from the entropy density derived from the ratio of the number of produced deuterons to nucleons. Some
differences between the numerical estimates and the
$\frac{T}{\varepsilon_f}$ expansions are found.

A method for measuring the temperature was proposed in \cite{15b}
based on momentum fluctuations of detected particles.  A
quadrupole $Q_{xy}=\langle p^2_x-p^2_y\rangle$ is defined in a
direction transverse to the beam axis (z-axis) and the average is
performed, for a given particle type, over events. Such a quantity
is zero in the center of mass of the equilibrated emitting source.
Its variance is given by the simple formula:
\begin{equation}
 \sigma^2_{xy}=\int d^3p(p^2_x-p^2_y)^2n(p)
\label{definequad}
\end{equation}
where n(p) is the momentum distribution of particles. In
\cite{15b} a classical Maxwell-Boltzmann distribution of particles
at temperature $T_{cl}$ was assumed which gives: $
\sigma^2_{xy}=\bar N 4m^2T_{cl}^2$,  m is the mass of the
fragment. $\bar N$ is the average number of particles.
 In heavy ion collisions,  the
produced particles  do  $\it not$  follow classical statistics
thus the correct distribution function must be used in Eq. (\ref{definequad}). Protons(p), neutrons(n), tritium etc. follow
the Fermi statistics while, deuterium, alpha etc., even though
they are constituted of nucleons, should follow the Bose
statistics. In this work we will concentrate on fermions only and
in particular p and n which are abundantly  produced in the
collisions  thus carrying important informations on the densities
and temperatures reached.
\begin{figure}
\centering
\includegraphics[width=1.0\columnwidth]{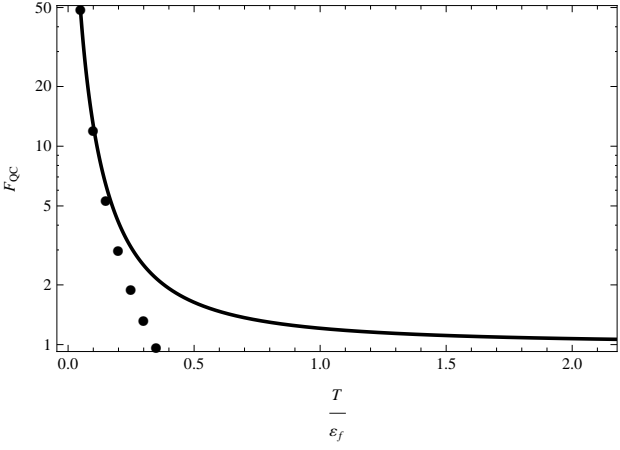}
\caption[]{$F_{QC}$ versus $\frac{T}{\varepsilon_f}$.  Symbols as
top panel in Fig .\ref{Fig1}.} \label{Fig2}
\end{figure}
Using a Fermi-Dirac distribution n(p)
\begin{eqnarray}
\langle \sigma_{xy}^2 \rangle&=&\frac{4}{15}\frac{\int_0^\infty dp p^6 \langle n_p \rangle}{\int_0^\infty dp p^2 \langle n_p \rangle}\nonumber\\
&=&\frac{4}{15}\frac{\frac{(2m)^\frac{7}{2}}{2}\int_0^\infty d\varepsilon \varepsilon^\frac{5}{2} \frac{1}{e^{(\varepsilon-\mu)/T}+1}}{\frac{(2m)^\frac{3}{2}}{2}\int_0^\infty d\varepsilon \varepsilon^\frac{1}{2} \frac{1}{e^{(\varepsilon-\mu)/T}+1}}\nonumber\\
&=&\frac{4}{15}\frac{\frac{(2mT)^\frac{7}{2}}{2}\int_0^\infty dy y^\frac{5}{2} \frac{1}{e^{y-\nu}+1}}{\frac{(2mT)^\frac{3}{2}}{2}\int_0^\infty dy y^\frac{1}{2} \frac{1}{e^{y-\nu}+1}}\nonumber\\
&=&(2mT)^2\frac{4}{15}\frac{\int_0^\infty dy y^\frac{5}{2} \frac{1}{e^{y-\nu}+1}}{\int_0^\infty dy y^\frac{1}{2} \frac{1}{e^{y-\nu}+1}}\nonumber\\
&=&(2mT)^2 F_{QC}(\nu) \label{numl_mom}
\end{eqnarray}
where $\varepsilon=\frac{p^2}{2m}$ is the energy , $\mu$ is the
chemical potential,  $\nu=\frac{\mu}{T}$.
$F_{QC}(\nu)$ is the quantum correction which
should converge to one for high T (classical limit).  Expanding to the lowest order in
$\frac{T}{\varepsilon_f}$, where
$\varepsilon_f=\varepsilon_{f0}(\frac{\rho}{\rho_0})^{2/3}=
36(\frac{\rho}{\rho_0})^{2/3}$ MeV is the Fermi energy of nuclear
matter,  the following result was obtained in  \cite{plb11,8}:
\begin{eqnarray}
\langle \sigma_{xy}^2
\rangle&=&(2mT)^2\frac{4}{35}(\frac{\varepsilon_f}{T})^2\times\nonumber\\
&&\left[1+\frac{7}{6}\pi^2(\frac{T}{\varepsilon_f})^2
+O(\frac{T}{\varepsilon_f})^4\right]\label{lowT_mom}
\end{eqnarray}
Within the same framework we can calculate the fluctuations of the p, n multiplicity distributions. These are given by \cite{8}:
\begin{eqnarray}
\frac{\langle(\Delta N)^2\rangle}{\langle N \rangle}
&=&\frac{\frac{gV}{h^3}4\pi\frac{(2mT)^\frac{3}{2}}{2}\int_0^\infty dy y^\frac{1}{2}\frac{e^{y-\nu}}{(e^{y-\nu}+1)^2}}{\frac{gV}{h^3}4\pi\frac{(2mT)^\frac{3}{2}}{2}\int_0^\infty dy y^\frac{1}{2}\frac{1}{e^{y-\nu}+1}}\nonumber\\
&=&\frac{\int_0^\infty dy
y^\frac{1}{2}\frac{e^{y-\nu}}{(e^{y-\nu}+1)^2}}{\int_0^\infty dy
y^\frac{1}{2}\frac{1}{e^{y-\nu}+1}} \label{numl_mul}
\end{eqnarray}
The lowest order expansion in  $(\frac{T}{\varepsilon_f})$, was also derived in \cite{plb11} and is given by:
\begin{equation}
\frac{\langle(\Delta N)^2\rangle}{\langle N
\rangle}=\frac{3}{2}\frac{T}{\varepsilon_f}
\label{lowT_mul}
\end{equation}

From the above equation (4) we can calculate numerically the multiplicity fluctuations for a given $\nu$ and recover the value of  $(\frac{T}{\varepsilon_f})$
from the following equation which is solved numerically:
\begin{equation}
\frac{T}{\varepsilon_f}=\frac{1}{\left[\frac{3}{2}\int_0^\infty dy y^\frac{1}{2}\frac{1}{e^{y-\nu}+1}\right]^{\frac{2}{3}}}\label{ratiotef}
\end{equation}
In Fig. \ref{Fig1} we plot the quantity $\frac{T}{\varepsilon_f}$
vs the normalized fluctuations obtained by solving numerically eqs.(4) and eqs.(6) while the lowest order approximation, eq.(5), is given by the full dots.
 Since in experiments or modeling
one recovers the normalized fluctuations, it is better to find a
relation between the normalized temperatures as function of the
normalized fluctuations displayed in the Fig. 1. It is useful to
parametrize the numerical results as:
\begin{eqnarray}
\frac{T}{\varepsilon_f}&=&-0.442+\frac{0.442}{(1-\frac{\langle(\Delta
N)^2\rangle}{\langle N \rangle})^{0.656}}\nonumber\\
&&+0.345\frac{\langle(\Delta N)^2\rangle}{\langle N
\rangle}-0.12(\frac{\langle(\Delta N)^2\rangle}{\langle N
\rangle})^2\label{fitfm}
\end{eqnarray}
 which is practically indistinguishable from the numerical result (full line) reported in Fig. \ref{Fig1}. 
As expected
the approximations contained in eq. (5) reproduce the numerical results (full line) up
to $\frac{T}{\varepsilon_f}\approx 0.5$.  
Since from experimental data or
models it is possible to extract directly the normalized
fluctuations, one can easily derive the value of
$\frac{T}{\varepsilon_f}$ from Eq. (\ref{fitfm}). 
 
 Before proceeding further,  it is important to test
the validity of the approximations for the quadrupole fluctuations
by comparing them to the numerical result solving Eq.
(\ref{numl_mom}).
\begin{figure}
\centering
\includegraphics[width=1.0\columnwidth]{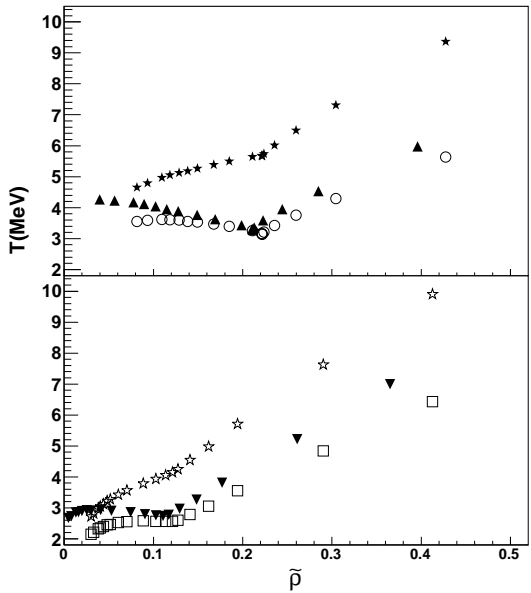}
\caption[]{Temperature versus density normalized to the ground
state density $\rho_0=0.16fm^{-3}$, derived from quantum fluctuations, Eqs.
(\ref{definequad}, \ref{numl_mom}, \ref{numl_mul}). Open dots and
open squares are the approximation at the lowest order in  $\frac{T}{\varepsilon_f}$ , full stars and open stars are the classical
cases similar to \cite{15b}, the full triangles are the numerical results. The top panel refers to
protons and the bottom panel refers to neutrons.} \label{Fig3}
\end{figure}
 In Fig. \ref{Fig2} we plot the quantum correction term $F_{QC}$ versus $\frac{T}{\varepsilon_f}$.
 The difference with the classical case is again striking (the $F_{QC}$ in Eq.(\ref{numl_mom}) equal to one for a classical perfect gas).
 For simplicity we can parametrize the numerical result with the simple approximation:
\begin{equation}
F_{QC}|_{fit}=0.2(\frac{T}{\varepsilon_f})^{-1.71}+1\label{fitquad}.
\end{equation}
which is indistinguishable from the numerical result displayed in Fig. \ref{Fig2}(full line). Clearly such
an equation converges to one at high T as expected.  Notice that the lowest order approximation \cite{plb11} is valid up
to a modest $\frac{T}{\varepsilon_f}\approx0.2$.
Eqs.(\ref{fitfm}) and (\ref{fitquad}) might be very useful when
deriving densities and temperatures from data or models, without
worrying if one is in the classical or fully quantum limit, the
only constraint is that we are dealing with fermions.

Once the density and the temperature of the system have been determined it is straightforward to
derive other thermodynamical quantities.  One of such quantities is the entropy:
\begin{equation}
S\equiv\frac{U-A}{T}=
N\left[\frac{5}{2}\frac{f_{5/2}(z)}{f_{3/2}(z)}-\ln
z\right]\label{entropy}
\end{equation}
where $f_m(z)=\frac{1}{\Gamma (m)}\int_0^\infty
\frac{x^{m-1}dx}{z^{-1}e^x+1}$ and $z=e^{\frac{\mu}{T}}$ is the
fugacity. U and A are the internal and Helmotz free energy
respectively\cite{8}.  This equation can be numerically evaluated
and the results are plotted in Fig. {\ref{Fig1}}(bottom panel).
 For practical purposes it might be useful to have a parametrization of the entropy in terms of the normalized fluctuations, which is physically transparent since entropy
 and fluctuations are strongly correlated \cite{8}:
 \begin{eqnarray}
 \frac{S}{N}|_{fit}&=&-41.68+\frac{41.68}{(1-\frac{\langle(\Delta N)^2\rangle}{\langle N \rangle})^{0.022}}\nonumber\\
 &&+2.37 \frac{\langle(\Delta N)^2\rangle}{\langle N \rangle}-0.83 (\frac{\langle(\Delta N)^2\rangle}{\langle N \rangle})^2\label{fitentropy}
 \end{eqnarray}
 The latter fit is indistinguishable from the numerical result plotted in Fig. {\ref{Fig1}} (full line-bottom panel) while the Sackur-Tetrod result (full triangles) is valid in the classical limit \cite{8} as confirmed in the figure 1.

To illustrate the strength of our approach we simulated
$^{40}Ca+^{40}Ca$ heavy ion collisions at fixed impact parameter
$b=1fm$ and beam energies  $E_{lab}/A$  ranging from 4 MeV/A up to
100 MeV/A.  Collisions were followed up to a maximum time $t=1000
fm/c$ in order to accumulate enough statistics.  Particles emitted
at later times (evaporation) could affect somehow the results and
this might be important especially at the lowest beam energies. A
complete discussion of these simulations can be found in
\cite{plb11}, here we will use the results to compare the different
approximations.

In Fig. \ref{Fig3} we plot the temperature vs density as obtained
from the quadrupole and multiplicity fluctuations. The top panel
refers to protons while the bottom to neutrons. As we can see from
the figure, the results obtained using the fit functions,
Eqs.(\ref{fitfm}) and (\ref{fitquad}), deviate slightly from the
lowest order approximations given in Eqs. (\ref{lowT_mom}) and
(\ref{lowT_mul}). This is a signature that we are in the fully
quantum regime for the events considered. For comparison, in the
same plot we display the classical temperatures which are
systematically higher than the quantum one, see
Eq.(\ref{numl_mom}) and Fig. \ref{Fig2} \cite{bauer}.  We notice that
for a given excitation energy we can derive a classical or a
quantum temperature, but the density can be derived for the
quantum case only within our approach. Of course other methods
could be devised that give both classical temperatures and
densities  using suitable fragment ratios \cite{albergo}.
We stress that those classical temperatures do not need to coincide
with the classical temperatures considered here since we are
dealing with protons and neutrons only.  Larger fragments could be
also included and a discussion on this can be found in
\cite{15,15b,15a}.\\
\begin{figure}
\centering
\includegraphics[width=1.0\columnwidth]{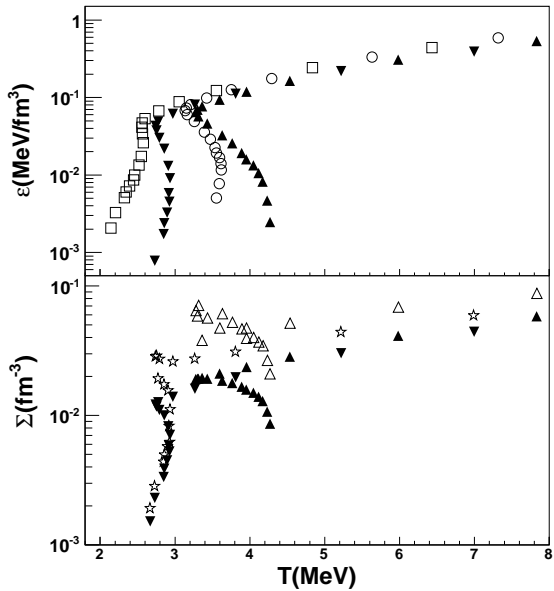}
\caption[]{(Top) Energy density versus temperature. Symbols as in Fig.
\ref{Fig3}; (Bottom)
entropy density versus temperature. The opens symbols refer to the entropy density calculated from the ratios of the produced number of deuterons to protons (triangles) (neutrons-stars), eq.(11).} \label{Fig4}
\end{figure}
To better summarize the results we plot in Fig. \ref{Fig4} (top panel) , the
energy density $\varepsilon=\langle \frac{E_{th}}{A}\rangle \rho$
versus temperature \cite{plb11}. Different particle types scale especially at
high T where Coulomb effects are expected to be small. A rapid
variation of the energy density is observed around $T\approx 2MeV$
for neutrons and  $T\approx 3MeV$ for protons which
indicates a first order phase transition \cite{michela}. As we see
from the figure, the numerical solution of the Fermi integrals
gives small corrections while keeping the relevant features
obtained in the lowest approximation intact.  This again suggests
that in the simulations the system is fully quantal.
We also notice that Coulomb effects become negligible at $T>>3MeV$ where the phase transition occurs. The smaller role of
the Coulomb field in the phase transition has recently been
discussed experimentally in the framework of the Landau's
description of phase transitions \cite{me}. 

 In order to confirm
the origin of the phase transition, it is useful to derive the
entropy density $\Sigma=\langle \frac{S}{N}\rangle \rho$ which is
plotted in the bottom panel of Fig. {\ref{Fig4}}. The rapid
increase of the entropy per unit volume is due to the sudden increase of the
number of degrees of freedom (fragments) with increasing T.
  The entropy can be also derived using the law of mass action from the ratio of the produced number of deuterons to protons (or neutrons) $R_{d,p(n)}$  \cite{Csernai,siemens}:
\begin{eqnarray}
 \frac{S}{N}|_{d/p(n)}&=&3.95-lnR_{d/p(n)}-1.25\frac{R_{d/p(n)}}{1+R_{d/p(n)}}\label{fitentropydn}
 \end{eqnarray}
The CoMD results from eq.(11)  multiplied by the density, are plotted in Figure 4 (bottom panel) with open symbols. We find an overall qualitative good agreement of the entropy density to  the quantum results, eq.10,
especially for neutrons.  Very interesting is the good agreement for neutrons at low T where the particles are emitted from the surface of the nuclei which is at low density, see also Fig.3.
Such a feature is not present for the protons due to larger Coulomb distortions.  There is a region near the transition ($T\approx 3MeV)$, where both ratios do not reproduce the quantum results.
However, at large temperatures it seems that all methods converge as expected.

In conclusion, in this work we have addressed a general method for
deriving densities and temperatures of fermions.
For high temperatures and small densities the
classical result is recovered as expected.
However, we have shown in CoMD calculations that the effect of
higher order terms give small differences in the physical
observables considered in this paper but they could become large when approaching the classical limit.
To overcome this problem we have produced suitable
parameterizations of quadrupole and multiplicity fluctuations
which are valid for fermions at all temperatures and densities. The results obtained in this paper are quite general and they could be applied
to other systems, for instance trapped Fermi gases \cite{prl},  to determine the entropy from normalized quantum fluctuations. We have also shown that the
quantum entropy can be compared to the one derived from the ratio of the number of deuterons to protons or neutrons produced in the collisions.  Especially the neutrons seem to give
cleaner results but of course they are more difficult to determine experimentally.

\end{document}